\documentclass[twocolumn,secnumarabic,amssymb, nobibnotes, aps, prb,superscriptaddress]{revtex4-1}

\usepackage{graphicx}
\usepackage{subfigure}

\usepackage{mathrsfs}
\usepackage{stmaryrd}
\usepackage{times}
\usepackage{amsmath}
\usepackage{amsfonts}
\usepackage{amssymb}
\usepackage{graphicx}
\usepackage{bm}
\usepackage{color}
\begin{document}
\makeatletter
\@ifundefined{textcolor}{}
{%
 \definecolor{CYAN}{cmyk}{1,0,0,0}
 \definecolor{MAGENTA}{cmyk}{0,1,0,0}
 \definecolor{YELLOW}{cmyk}{0,0,1,0}
 }

\title{Quench Dynamics of Josephson Current in a Topological Josephson junction}
\author{Dihao Sun}
\affiliation{ Department of Applied Physics, School of Science, Xi'an Jiaotong University, Xi'an 710049, China}
\author{Jie Liu}
\email{jieliuphy@xjtu.edu.cn}
\affiliation{ Department of Applied Physics, School of Science, Xi'an Jiaotong University, Xi'an 710049, China}

\begin{abstract}
 The $4\pi$-periodic Josephson Effect is a distinguishing feature of a topological Josephson junction. However, stringent conditions make it hard to observe in experiments. In this work, we study the transient transport properties in a topological Josephson junction  numerically. We show that the $4\pi$ Josephson current can be sustained  under nonequilibrium conditions. The properties of the Josephson current are analyzed for different conditions, and three main regimes are identified: First, when both the superconducting wires of the Josephson junction lie in the topologically nontrivial region, a $4\pi$ Josephson current can appear upon suddenly applying a DC voltage. Second, when one superconducting wire lies in the trivial region, while the other wire lies in the non-trivial region, the Josephson current is $2\pi$ periodic but the component of the higher order Josephson current increases. Third, when both wires lie in the trivial region, a stable $2\pi$ Josephson current is observed.
 Most importantly, the fractional Josephson Effect is fragile
 in the presence of disorder. Hence, experiments should be designed  carefully to eliminate the effect of disorder. These results could be helpful to optimize fine-tuning of the experimental parameters to observe the $4\pi$-periodic Josephson current in a topological Josephson junction.
\end{abstract}

\maketitle

\section{Introduction}

Due to the presence of Majorana quasi-particle states (MQP) at their ends, topological superconductors are viewed as the most promising platform for fault-tolerant quantum computation \cite{kitaev,nayak}. Various strategies have been proposed for the realization of topological superconductors \cite{Fu,sau,fujimoto,sato,alicea2,lut,oreg,potter,S1,Ki2}. Among those proposals, a semiconductor wire, which is subject to an external magnetic field with Rashba spin-orbit coupling and proximity-induced superconductivity, has been singled out as the most feasible device \cite{sau}. Indeed, due to recent advances in state of the art nanotechnology, more and more groups have reported the fabrication of topological superconductor systems based on semiconductor wires, and the detection of the MQP signal in such systems\cite{kou,deng,das1,perge}. Recently, Kouwenhoven et al. have further improved the fabrication technology and constructed a ballistic semiconductor superconducting wire. Such a highly clean system can eliminate the influence of disorder and makes the results more plausible \cite{hao}. In addition, Marcus et al. have demonstrated the existence of MQPs through their teleportation property \cite{Marcus}.  Perge et al. have also provided a high resolution signal of MQPs in an atomic chain topological superconductor system \cite{Yaz2}. While these results certainly demonstrate the arrival of MQPs, they are mostly related to the local density of states of MQPs. Conversely, the amount of available results concerning other unique properties of MQPs is still very scarce, which raises the need for further studies in this field.

    Another hallmark of topological superconductors is the putative fractional Josephson effect (FJE) \cite{law1,Fu1,Jiang}. When two topological superconductor wires are put together, a topological Josephson junction (TJJ) is formed, which can support single electron tunneling with a period of $4\pi$. However, this is different from a conventional Josephson junction, which allows tunneling by Cooper pairs only, with a period of $2\pi$. Since the $4\pi$ Josephson effect is a unique transport property of MQPs, a lot of research has been dedicated towards its realization.
 Several groups have attempted to build a superconductor-topological insulator-superconductor junction which is also expected to present
 the $4\pi$ Josephson current \cite{yaco,kou2,lu,Erw1, Erw2,Deacon3}. Indeed, Molenkamp's group has {\color{CYAN} showed the signal of} $4\pi$-periodic Josephson Effect in such a junction \cite{Erw1, Erw2,Deacon3}. However, the fractional Josephson effect in a semiconductor superconducting wire system has not been experimentally observed yet.
  Kouwenhoven  and Marcus et al. succeeded at fabricating this type of junction. However, they did not observe the $4\pi$ periodicity \cite{kou1,nkou1,nkou2,nMar1}. There are two significant experimental challenges in semiconductor superconducting wire systems:
first,  this type of Josephson junction is composed of two semiconductor superconducting wires, both of which should lie in the nontrivial region, thus making its realization a doubly difficult process. Second, the presence of a $4\pi$ periodicity requires parity conservation, which is a
stringent condition \cite{pablo}.
 As shown in Fig. 1(a), in addition to the pair of MQPs located at the junction, there is an additional pair of MQPs localized at its exterior ends. These two pairs of MQPs may hybridize with each other, destroying the parity even though the hybridization strength would decrease exponentially with the length of the wire.

     In order to overcome the first obstacle, a better understanding of the Josephson current properties of the TJJ is required,
     especially as far as its
      behavior in different regions is concerned. This can allow identification of the region where the system lies in, in order to fine-tune both wires into the nontrivial region. With respect to the second obstacle, previous researches indicate that that the FJE can be recovered in a non-equilibrium situation on a finite time scale only\cite{pablo,Plat1,Fran,virt,houzet,shu}, because it inevitably decays to $2\pi$ period over time. To reveal the $4\pi$ information in a long time scale, some indirect means may be used, such as the even-odd Shapiro step or noise measurements \cite{Jay1,Erw4,Plat2}. Several experiments have indeed led to the observation of the even-odd Shapiro steps \cite{Frac}. However, these signals are indirect and easily affected by the environment. It is therefore preferable to attempt direct visualization of the $4\pi$ Josephson current.

        In this paper, we study the AC Josephson current in a TJJ under a suddenly applied DC voltage. {\color{CYAN} Motivated by recent advance in experiment \cite{Deacon3}, we study the regimes of a constant driving in a semiconductor superconducting wire Josephson junction. }
        We compare the Josephson current under several different conditions: first, we demonstrate that a $4\pi$-periodic Josephson current can appear if both topological superconductor wires lie in the nontrivial region. If one wire lies in the trivial region and the other one lies in the nontrivial region, an unstable $2\pi$ periodic Josephson current is observed. Finally, if both topological superconducting wires lie in the trivial region, a stable $2\pi$ Josephson current is observed. Identifying the current behavior in these different situations can facilitate the experimental detection of a real $4\pi$ Josephson current. Furthermore, we show that such $4\pi$ Josephson current is very sensitive to disorder. Although MQPs are robust against disorder, the background is sensitive to disorder, hence the fractional Josephson effect could be completely destroyed even in the presence of very weak disorder.

 \begin{figure}[ptb]
\includegraphics[width=1.0\columnwidth]{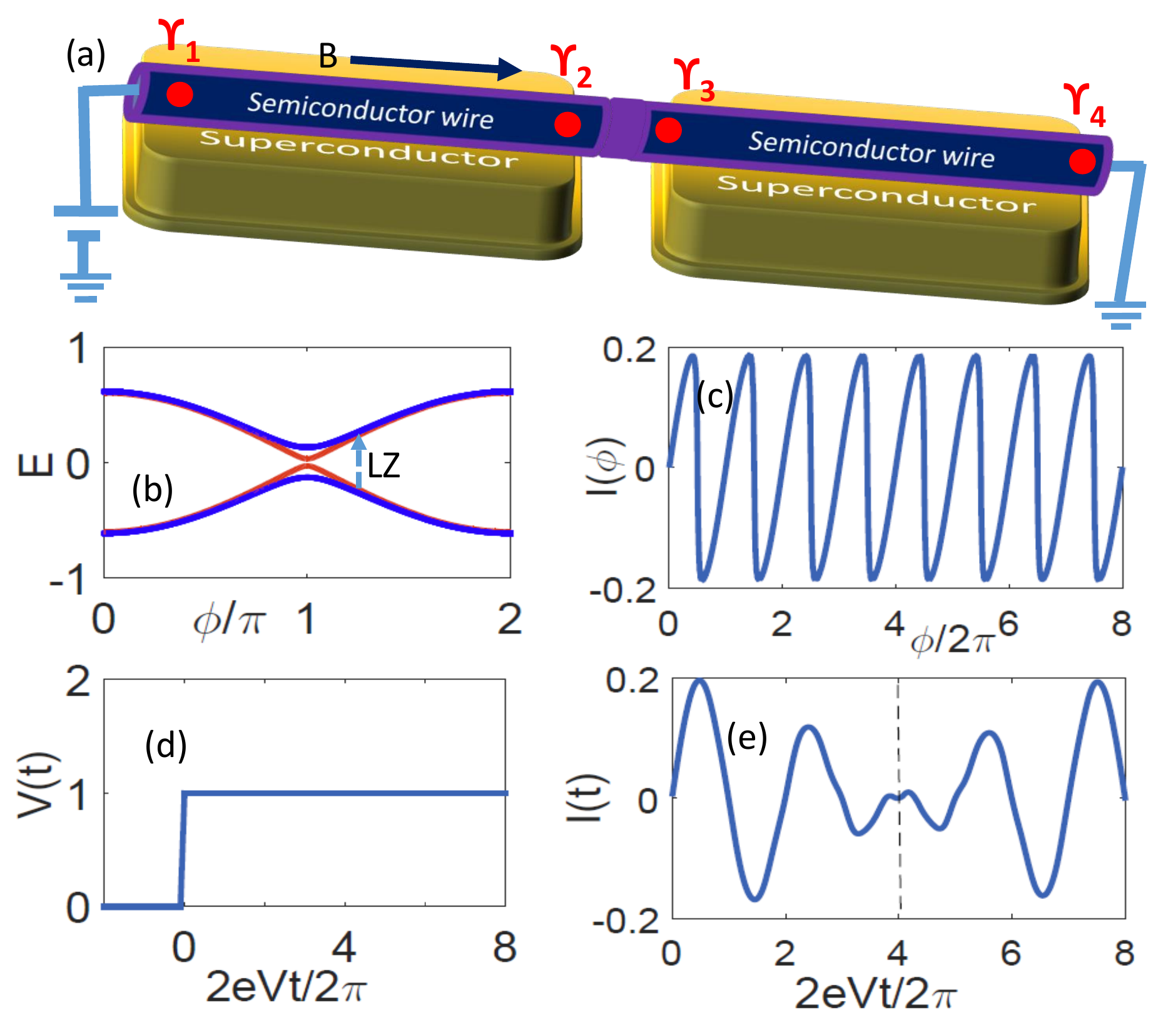}
\caption{(Color online) (a) Schematic diagram for topological superconductor-insulator-topological superconductor Josephson junction. Four MQPs exist at the ends of the wire when the system lies in topological nontrivial region. (b) The energy spectrum of Andreev bound states in topological nontrivial case, as a function of SC phase difference $\phi$. Due to the finite wire length, MQPs would hybridize with each other and open a gap.(c) Supercurrent in the adiabatic regime is $2\pi$ periodic.(d) A step voltage $V(t)=V\theta(t)$ used to induce nonadiabatic $4\pi$ periodic current. (e) The supercurrent induced by  the step voltage of (d). Here $\phi=2eVt$ which is driven by the step voltage. We can see that the period is $4\pi$ and twice the period of (c), although the evolution of current may switch suddenly at the position indicated by the vertical dashed line.}
\label{Figure1}
\end{figure}

The rest of this paper is organized as follows: In Sec.~\ref{sec:models}, the Hamiltonian of TJJ and the formula to calculate supercurrent is introduced. In Sec.~\ref{sec:discussions}, we  show the numerical results into three parts: In Subsec.~\ref{sec:1}, the properties of the Josephson current in nontrivial case are shown. In Subsec.~\ref{sec:2}, we compare the Josephson current in different conditions, including both the trivial and the nontrivial cases.  In Subsec.~\ref{sec:3}, we show that the Josephson current is very sensitive to disorder. Finally, a brief summary is given in Sec.~\ref{sec:conclusions}.

\section{Model and Formalism}
\label{sec:models}

A typical TJJ is composed of two topological superconducting wires with different superconducting phases.
Following  Refs.[\onlinecite{jie1,jie2}], the tight-binding model Hamiltonian of the superconducting wire can be written as:
\begin{eqnarray}\label{model1}
 H_{s,q1D}& =& \sum\nolimits_{\mathbf{i},\mathbf{d},\alpha } { -t_0(\psi _{\mathbf{i} + \mathbf{d},\alpha }^\dag  \psi _{\mathbf{i},\alpha }  + h.c.) - \mu_s \psi _{\mathbf{i},\alpha }^\dag  \psi _{\mathbf{i},\alpha } } \nonumber \\
&-& \sum\nolimits_{\mathbf{i},\mathbf{d},\alpha ,\beta } { i{U _{\mathbf{R}}} \psi _{\mathbf{i} + \mathbf{d},\alpha }^\dag  \hat z \cdot (\vec{\sigma}  \times \mathbf{d})_{\alpha \beta }   \psi _{\mathbf{i},\beta } } \nonumber  \\
 &+& \sum\nolimits_{\mathbf{i},\alpha ,\beta } { \psi _{\mathbf{i}, \alpha }^\dag [(V_x \sigma_x)_{\alpha \beta} +V_{\text{imp}}(\mathbf{i})\delta_{\alpha \beta}]\psi _{\mathbf{i},\beta } } \nonumber \\
& +& \sum\nolimits_{\mathbf{i},\alpha} \Delta e^{i\phi_s} \psi _{\mathbf{i},\alpha }^{\dagger} \psi _{\mathbf{i},-\alpha }^{\dagger}+h.c. \\
H_{c} & = & \sum\nolimits_{\alpha }(t_c \psi _{LN,\alpha}^{\dagger} \psi _{R1,\alpha}+h.c.).
 \end{eqnarray}
Here, $H_{s,q1D}$ is the Hamiltonian of the left (right) wire with $s=L$ ($R$). $\mu_{L (R)}$ means the chemical potential in the left (right) wire which can be tuned independently and may thus have different values for each wire. In addition, the phases of the superconducting order, $\Delta e^{i\phi_s}$ are also different for the two wires (here we set $\phi_L=\phi$ and $\phi_R=0$). All other parameters are the same for both wires. Furthermore, $\mathbf{i}$ denotes the lattice site, and $\mathbf{d}$ denotes the unit vector which connect the nearest neighbor sites in the $x$ directions. $\alpha, \beta$ are the spin indices. $t_0$ is the hopping amplitude, $U_{R}$ is the Rashba coupling strength, and $V_{x}$ is the Zeeman energy caused by the magnetic field along the wire direction. $\Delta$ is the superconducting pairing amplitude and $V_{\text{imp}}(\mathbf{i})$ is  the on-site impurity.
$H_{c}$ describes the Josephson coupling between the left topological superconducting wire and the right one.

To study the FJE in the TJJ system, a step voltage of the form $V(t)= V\theta(t)$ is considered, as shown in Fig. 1(d). The voltage is zero when $t<0$ and $V$ when $t>0$. After applying this DC voltage to the left wire, the chemical potential of left wire turns into $\tilde{\mu}_L(t)=\mu_L+V\theta(t)$. In addition, the superconducting order parameter in the left wire depends on the applied voltage as $\Delta_L=\Delta e^{i2e/h\int_0^tV(t)dt}$. By performing a unitary transformation $U(t) = exp[\sum\nolimits_{\mathbf{i},\alpha }i(\phi/2+\frac{e}{h}\int_0^tV(t)dt)\psi _{\mathbf{i},\alpha }^\dag  \psi _{\mathbf{i},\alpha }]$, $\tilde{\mu}_L$ will transform to $\mu_L$ and
$\Delta_L=\Delta$. Thus,  the Hamiltonian of the whole system can be written as follows:
\begin{eqnarray}\label{model2}
H(t)&=& H_{L,q1D}(\phi=0)+H_{R,q1D}+H_{c}(t), \nonumber\\
H_{c}(t) &=& \sum\nolimits_{\alpha }(t_ce ^{i(\phi/2+\frac{e}{h}\int_0^tV(t)dt)} \psi _{LN,\alpha}^{\dagger} \psi _{R1,\alpha}+h.c.).
 \end{eqnarray}
 Where the only time dependent term is found in the Josephson coupling $H_{c}(t)$.

To calculate the supercurrent at the junction, we use the following current definition \cite{caio,Lei}: $\hat{J}_{\mathbf{i}}=-\frac{i}{2}\sum_{\mathbf{j},\alpha,\beta}\delta_{\mathbf{ij}}(t_{\mathbf{i}\alpha,\mathbf{j}\beta}\psi^{\dagger}_{\mathbf{i}\alpha}\psi_{\mathbf{j}\beta})$. Here, $\hat{J}_{\mathbf{i}}$ means the current flowing through site ${\mathbf{i}}$ (in this system the site locates at the junction), $\delta_{\mathbf{ij}}$ is the vector displacement of site $\mathbf{i}$ from site $\mathbf{j}$, and  $t_{\mathbf{i}\alpha,\mathbf{j}\beta}$ is the hopping parameter from site $\mathbf{i}$ with spin $\alpha$ to site $\mathbf{j}$ with spin $\beta$.
 Through the current definition, the time dependent current can be calculated as
 \begin{equation}\label{equation}
 I(t)=\langle \hat{J}_{\mathbf{LN}} \rangle = -\frac{i}{2}(\langle \psi_{LN}(t)|H_c(t)|\psi_{R1}(t) \rangle -h.c.).
 \end{equation}
 Here, $|\psi_{LN}(t)\rangle$ is the wave function at the right end of the left wire at time $t$, and $|\psi_{R1}(t) \rangle$ is the wave function at the left end of the right wire at time $t$. The time dependent wave function can be calculated by using the general time-evolution function $|\psi(t) \rangle= \sum_{n}exp(-i\int_0^t H(t)dt)|\psi_n(0) \rangle$. Here, $|\psi_n(0) \rangle$ is the $n$th initial wave function of $H(t=0)$.

\section{Results and Discussion}~\label{sec:discussions}
~\label{sec:discussions}

In this section we calculate the transient current by applying Eq. (\ref{equation}) in different situations. To facilitate the study, the superconducting pairing amplitude $\Delta=250\mu eV$ is set as the unit value. The other parameters are set as follows: $t_0=10\Delta, V_x=2\Delta, U _{\mathbf{R}}=2\Delta$.

Due to the finite length of the wire, the two MQPs localized at the ends of the wire hybridize with each other. The effective hybridization strength of the MQPs is of the order of $E_M=e^{-L/\xi}$. Here, $L$ is the length of the wire and $\xi=t/\Delta$ is the superconducting coherence length.
In this situation, the effective low energy Hamiltonian of the TJJ can be described by the following equation: $H_{eff}=iJ_1cos(\phi/2)\gamma_2\gamma_3+iE_M\gamma_1\gamma_2+iE_M\gamma_3\gamma_4$. Here, $J_1$ is the Josephson coupling between the two MQPs at the junction, and $E_M$ indicates the hybridization strength of the two MQPs in each wire as shown in Fig. 1(a). We set them as equal for simplicity. When $E_M$ is strictly equal to zero, the energy spectra of Andreev bound states are $E=\pm J_1cos(\phi/2)$. The spectra intersect at $\phi=\pi$ as shown by the red line in Fig. 1(b). In this case, the Josephson current is strictly $4\pi$ periodic, while when $E_M$ is not zero, the energy spectra of the Andreev bound states open a gap at $\phi=\pi$ as shown by the blue line in Fig. 1(b). In this case, the energy spectrum evolves adiabatically along the $E<0$ spectrum even though $E_M$ is exponentially small. Fig. 1(c) shows the Josephson current in the adiabatic evolution process, which decays to a $2\pi$ periodic form. However, the $4\pi$ periodicity can be recovered in a nonequilibrium situation via a non-adiabatic process like Landau-Zener transition as indicated by the dashed arrow in Fig. 1(b) \cite{pablo,Plat1,Fran,virt,houzet,Jay1}. Fig. 1(e) shows the typical transient Josephson current induced by the step voltage as shown in Fig. 1(d). The calculated current is $4\pi$ periodic, although
 may easily switch a $\pi$ phase as indicated by the dashed line in Fig. 1(e).  This $\pi$ phase switching is caused by the Landau-Zener-Stuckelberg interference effect \cite{pablo,Plat1,Fran,virt,houzet,Wang}.

\begin{figure}[ptb]
\includegraphics[width=1.0\columnwidth]{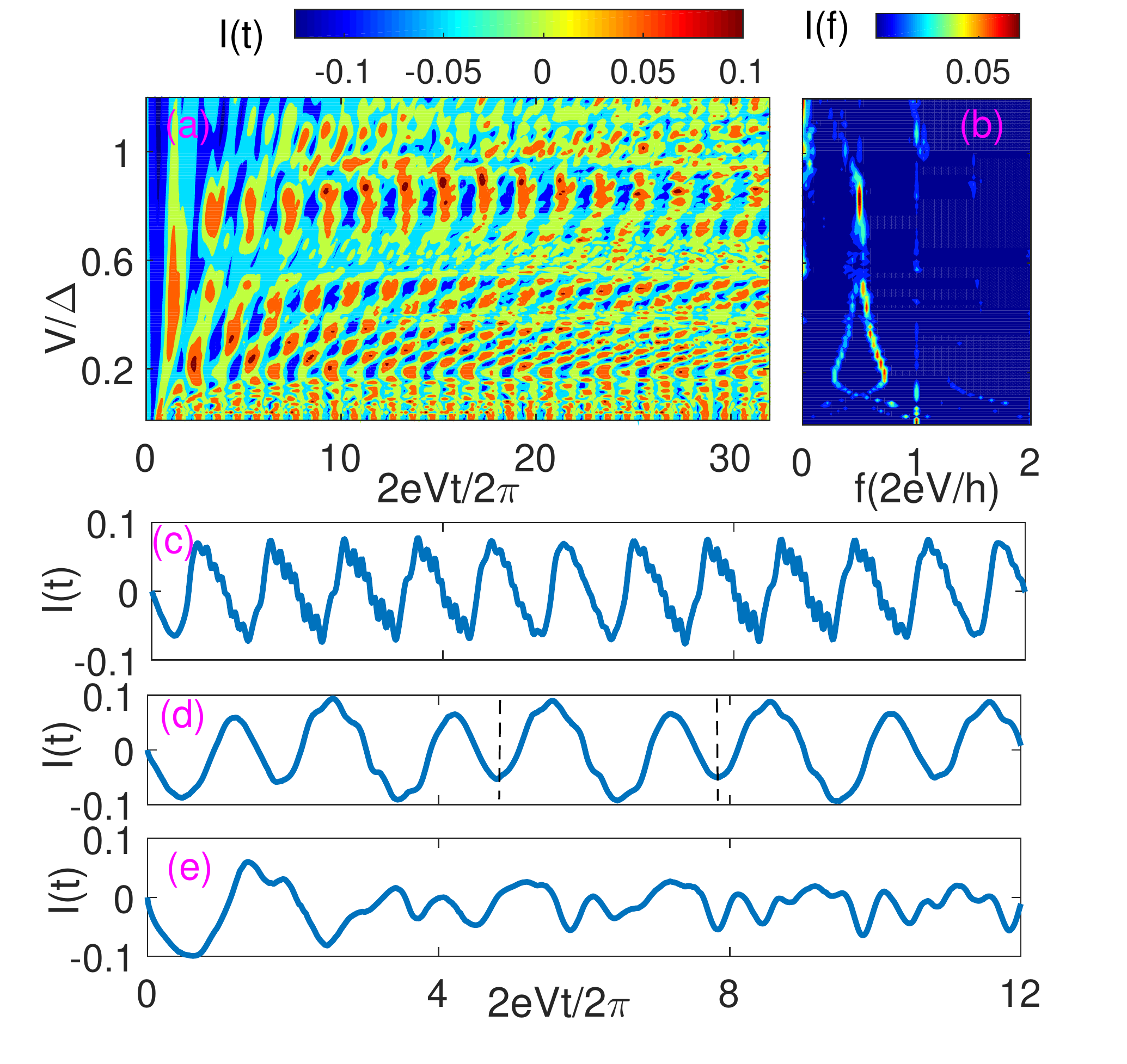}
\caption{(Color online) (a)Transient current versus the voltage and time period. Here, the length of the superconducting wire is about twice of coherence length, and $E_M=0.1\Delta$. (b) The Fourier transformation of transient current.  The current can be divided into three distinct regions: (c) Typical $2\pi$ periodic supercurrent in adiabatic region with $V<E_M$. The plot is extracted from the contour plot at $V=0.02\Delta$. (d) $4\pi$ current region with $E_M<V<\Delta_{eff}-J_1$. The plot extracts from Fig. 2(a) at $V=0.3\Delta$. (e) Dissipative current region with $V>\Delta_{eff}-J_1$. The plot shows the transient current at $V=0.75\Delta$, the average current over a period deviates from zero. }\label{Figure2}
\end{figure}

\subsection{Josephson current in the topological nontrivial region} ~\label{sec:1}

As a following step, we study the properties  of the Josephson current in the topological nontrivial region using the tight binding model according to Eq. (3). In the following we will use $2eVt$ instead of $\phi$ because $\phi=2eVt$ varies with time. In order to analyze the effects of the finite length, we initially study a short wire case. Fig. 2(a) shows the contour plot of supercurrent versus voltage and time period with $\mu_L=\mu_R = -2t_0$ and $t_c=0.6t_0$. Here $N_L=N_R=20a$, the wire length is about twice as much as the coherence length. In this situation, the effective parameters are: $J_1\approx 0.3\Delta$, $E_M\approx 0.1\Delta$. In addition, the effective superconducting gap $\Delta_{eff}\approx 0.85\Delta$ is suppressed by magnetic field.
The properties of the Josephson current can be divided into three distinct regions as a function of voltage $V$. First, the adiabatic region is found when $V<E_M$. Fig. 2(c)
is the typical plot extracted from the Fig. 2(a)  at $V=0.05\Delta$. In this regime, the Josephson current is approximately $2\pi$ periodic since the applied voltage can not support the Landau-Zener transition. However, some differences to the real adiabatic $2\pi$ current still emerge in Fig. 2(c). The Josephson current in Fig. 2(c) displays a fast oscillation due to the sudden voltage transition. In addition, a sudden $\pi$ phase switching occurs during the evolution of the process. When $E_M<V<\Delta_{eff}-J_1$,  the plot in Fig. 2(d) which is extracted from Fig. 2(a)  at $V=0.3\Delta$ shows different periods. In this case,  the voltage is high enough to support the Landau-Zener transition, and the $4\pi$ periodic Josephson current can be recovered.
As shown in Fig. 2(d), the period is twice as long as the period observed in Fig. 2(c). Although $\pi$ phase switching process occurs as indicated by the vertical dashed line, the $4\pi$ Josephson current can be sustained for a significant amount of time. The third region corresponds to a higher applied voltage, i.e. $V>\Delta_{eff}-J_1$. Fig. 2(e), which is extracted from the Fig. 2(a) at $V=0.75\Delta$, shows that the initial period is still $4\pi$. However, this condition only lasts for a short time. Eventually, additional peaks appear, and the average current deviates from zero after the initial period. This current is in fact a flow of electrons through the continuous band when $V>\Delta_{eff}$.

To extract the information contained in the transient current, Fig. 2(b) shows the Fourier transformation of Fig. 2(a). The amount of the time period used for Fourier transformation is $32$ oscillation periods here and in the following figures. We should stress here that the frequency does not depend on the amount of time used for Fourier transformation, although the amplitude of frequency fluctuates dramatically with the variation of time period. The three regions are distinct on the frequency map.  When $V<E_M$,  the frequencies are mainly distributed around $f_0=2eV/h$, which indicates the trivial Josephson Effect. On the other hand, when $E_M<V<\Delta_{eff}-J_1$, the frequencies are mainly distributed around $f_0/2 = eV/h$. This is half the frequency of the trivial Josephson current and therefore the fractional Josephson Effect occurs. The third region is corresponds to the dissipative current region,  where the net current is non-zero. Thus, the amplitude of zero frequency becomes large. Moreover, the frequency map suggests that the Landau-Zener transition assisted current can be described by a typical beating effect: $I(t) = I_1cos((f_0/2+\delta)t)+I_2cos((f_0/2-\delta)t)$. Here $I_1$ and $I_2$ are the amplitudes of the current, and $\delta$ is the frequency shift,  which is
determined by the relation between V and $E_M$, where the frequency shift $\delta  \rightarrow 0$ when $V\gg E_M$, and $\delta \rightarrow f_0/2$ when $V\ll E_M$. Thus,  the transition from the adiabatic region to the Landau-Zener region can be completely described by the beating effect.
\begin{figure}[ptb]
\includegraphics[width=1.0\columnwidth]{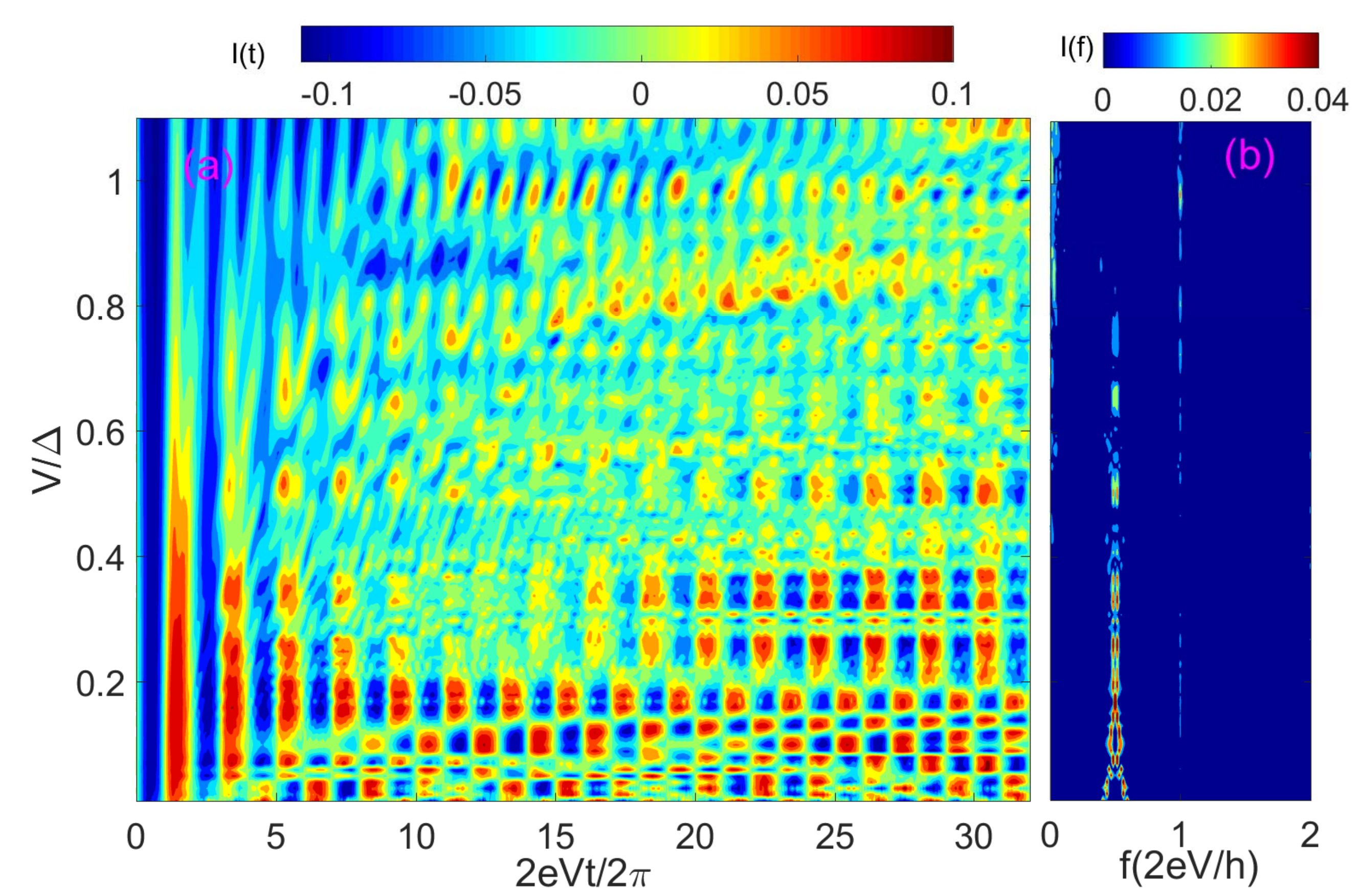}
\caption{(Color online)(a)  Transient current versus voltage and time period. Here, the length of wire is five times of the coherence length, and $E_M=0.01\Delta$. (b) The Fourier transformation of transient current. The frequency map shows clear fractional frequency signal at $eV/h$ When $V<\Delta_{eff}-J_1$. } \label{Figure3}
\end{figure}

The three regions with distinct properties of the Josephson current identified above become less distinct due to the large hybridization strength of the MQPs. To learn more about the $4\pi$ Josephson current, it is necessary to increase the wire length. Fig. 3 shows the transient current flowing through a wire whose length is five times the superconducting coherence length that can be achieved in an experiment typically  \cite{kou}. Here, $N_L=N_R=50a$, $E_M\approx0.01\Delta$. Due to the small hybridization strength of MQPs, only two distinct regions appear in Fig.3(a). The first region represents the $4\pi$ periodic Josephson Effect, while the other one is the dissipative current region. The $4\pi$ periodic Josephson current region can be seen more clearly than for the short wire case. Fig. 3(b) shows the Fourier transformation of Fig. 3(a).  We can see that the frequency is mainly distributed around $f_0/2 = eV/h$ when $V<0.6\Delta$.  However, when $V>0.6\Delta$, the weight of the zero frequency increases with increasing voltage. This corresponds to the dissipative current region with $V>\Delta_{eff}-J_1$, as discussed above.
\begin{figure}[ptb]
\includegraphics[width=1.0\columnwidth]{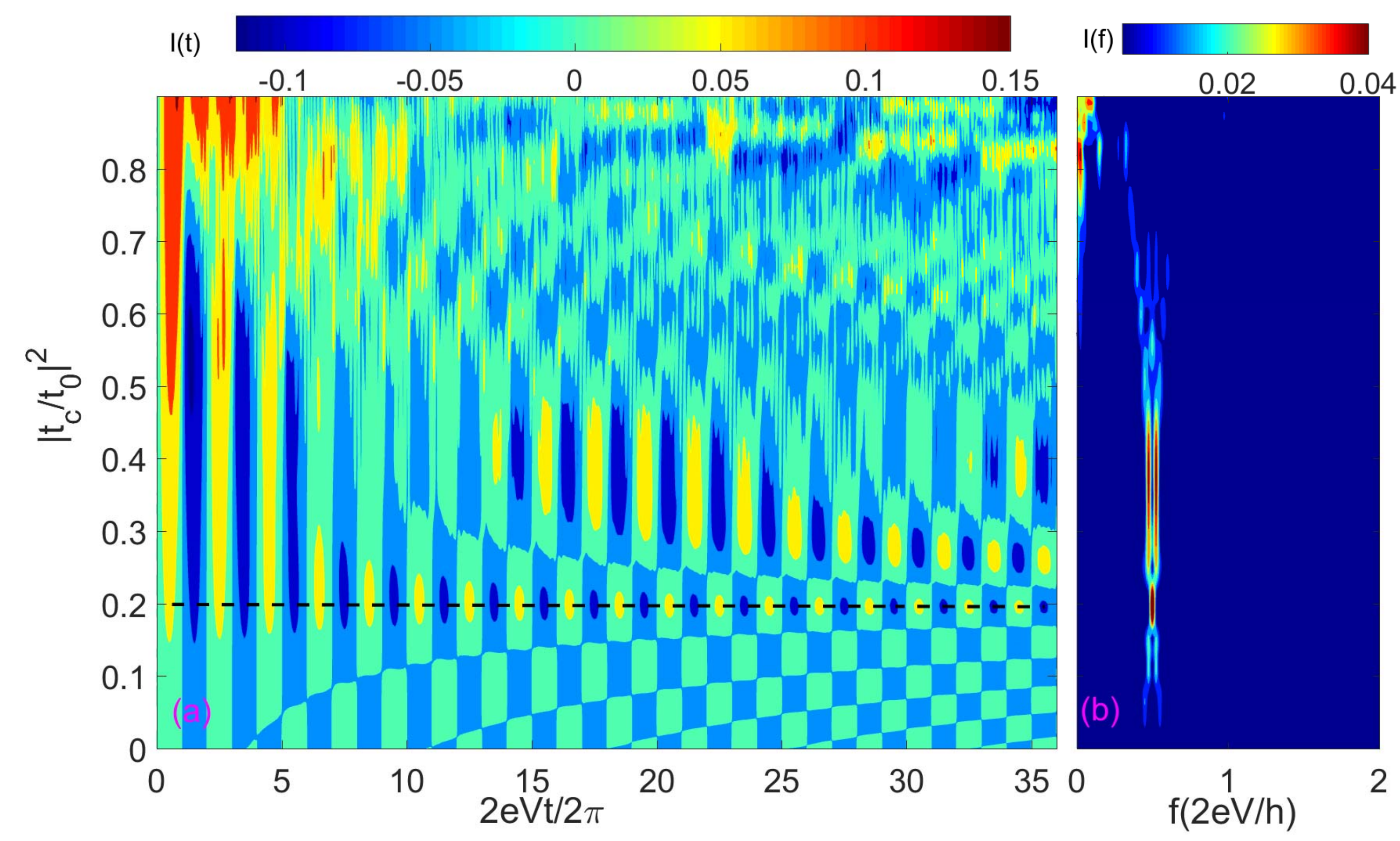}
\caption{(Color online) Contour plot of transient current versus the coupling strength and time period. Two semiconductor superconducting wires connect perfectly with each other when $t_c/t_0$ is 1. In contrast, the two semiconductor superconducting wires are disconnect when $t_c/t_0$ is 0. The $4\pi$ Josephson current is easy to destroy in strong-link case, while it can sustain a long time in weak-link case. (b) The Fourier transformation of transient current. Here, $N_L=N_R=50a$,$\mu_L=\mu_R=-2t_0$.}\label{Figure4}%
\end{figure}

 We have shown that the transient current features a $4\pi$ periodic Josephson component in the nontrivial region when $E_M<V<\Delta_{eff}-J_1$. Next, we investigate the effect of Josephson coupling $J_1$.  When $J_1+V>\Delta_{eff}$, the states in the Andreev bound states can jump to the bulk band with the help of voltage. In this case, the $4\pi$ Josephson current could be easily spoiled by the bulk band states \cite{pablo}. In our previous studies, we set $t_c=0.6t_0$, which leads to $J_1\approx0.3\Delta$. In the following, we adjust $t_c$ to change the effective Josephson coupling. As shown in Ref. [\onlinecite{Fu1}], $J_1=\sqrt{D}\Delta_{eff}$. Here $D$ is the transmission probability of the junction which varies monotonously with the coupling strength $t_c$ at the junction. If $|t_c/t_0|^2$ approaches 1, then $D$ approaches 1; while if $|t_c/t_0|^2$ approaches 0, then $D$ approaches 0.  Fig. 4(a) shows the contour plot of the transient current $I(t)$ as the function of the square of the coupling strength $|t_c/t_0|^2$, with $V=0.1\Delta$. Fig. 4(b) shows the corresponding Fourier transformation.
   As $|t_c/t_0|^2$ approaches 1, the stability of the $4\pi$ periodic current decays rapidly.
    This is consistent with Ref. [\onlinecite{pablo}], where the stability of the $4\pi$ periodic supercurrent is strongly influenced by the distance between the bulk band and the Andreev bounds state formed by MQPs. As $|t_c/t_0|^2$ approaches 1, the Josephson coupling $J_1$ approaches  $\Delta_{eff}$. In this case,  the Andreev bound states are easily influenced by the bulk band due to the small distance between the Josephson coupling and bulk band. Therefore, the $4\pi$ periodicity vanishes  rapidly when the connection is perfect. The most important conditions for the observation of the FJE can thus be summarized as follows: the length of the wire should be long enough and the connection at the junction should be weak.
   Another interesting result is that the $4\pi$ transient current can be sustained for a long time without $\pi$ phase switching at $|t_c/t_0|^2=0.2$ as indicated by dashed line. In this situation, the effective Josephson coupling strength is $J_1\approx 2V=0.2\Delta$. This indicates the presence of a stable $4\pi$ periodic Josephson current at $J_1\approx2V$. Fig. 3(a) confirms that a durable $4\pi$ periodic current without $\pi$ phase switching is observed at $J_1\approx 2V \approx0.3\Delta$. Such conditions obtained by numerical studies are expected to be useful for further manipulation of MQPs.

  \begin{figure}[ptb]
\includegraphics[width=1.0\columnwidth]{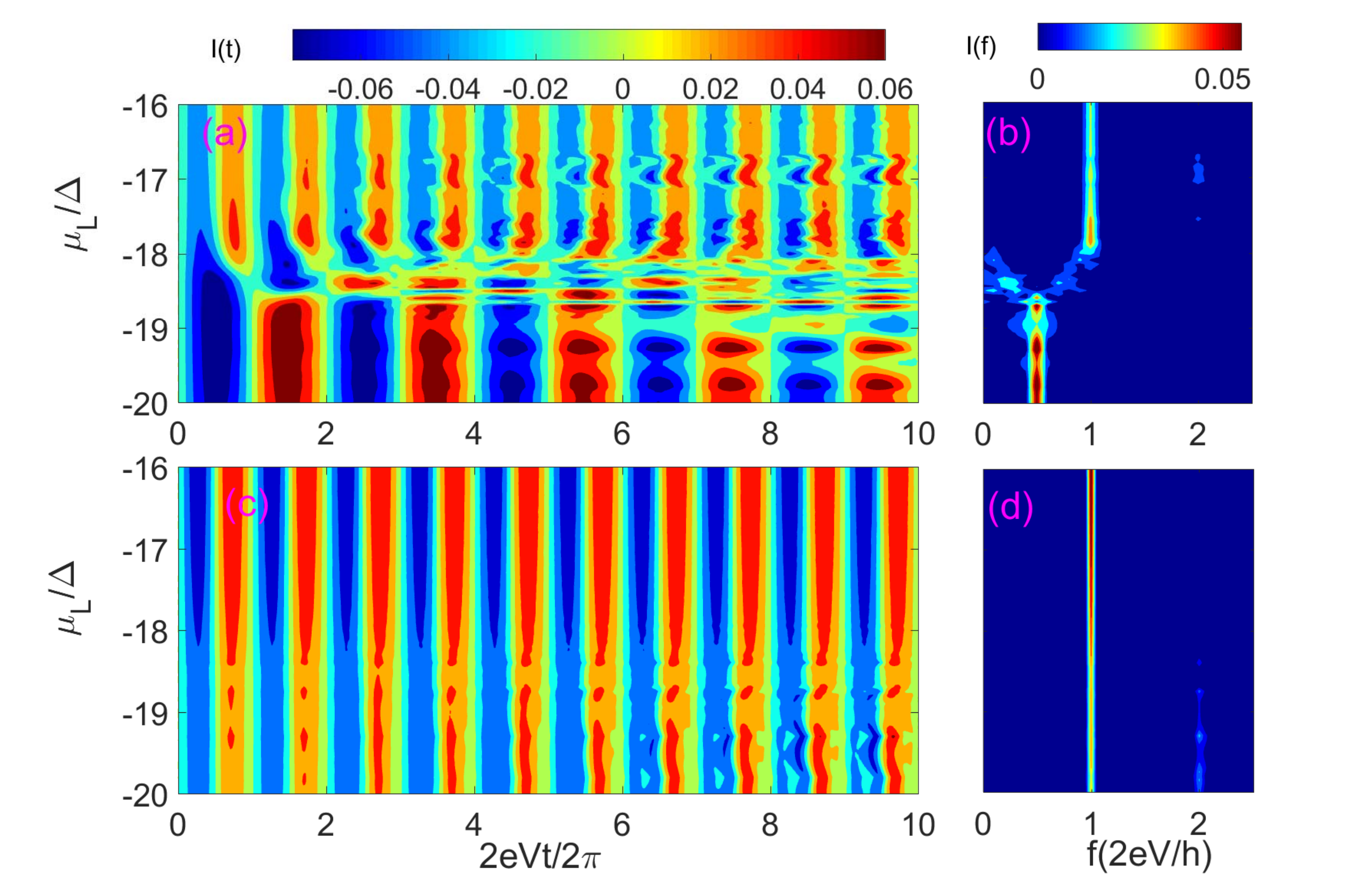}
\caption{(Color online) The behavior of transient current in different chemical potential with fixed voltage $V=0.1\Delta$. (a) The contour plot versus the chemical potential $\mu_L$ and time period with $\mu_R = -2t_0$ keeping in the non-trivial region. The $4\pi$ periodic current is clear when both wires lie in the non-trivial region, while the Josephson current is $2\pi$ periodic when the left wire lies in the trivial region. (b) Fourier transformation of (a).  (c)  The contour plot versus time period and the variance of $\mu_L$ with $\mu_R = -2t_0+3\Delta$ keeping in the trivial region. We can see that the period is always $2\pi$ periodic. However, the $2\pi$ Josephson current is unstable when one wire lies in the nontrivial region.  (d) Fourier transformation of (c).}\label{Figure5}%
\end{figure}

\subsection{Comparison  of the Josephson current for different regions}  ~\label{sec:2}

The $4\pi$ periodicity can occur under suitable conditions in the nontrivial region.
However, compared to the trivial region, the nontrivial region only takes up a small portion.
Tuning the system into the nontrivial region requires delicately gate the chemical potential. In our previous studies we assumed $\mu_L=\mu_R$ for simplicity. However, the chemical potentials are not generally the same in both wires. Three scenarios can occur: First, both wires lie
in the nontrivial region, which corresponds to the previous $4\pi$ Josephson current region; Second, both wires lie in the trivial region; Third, one wire lies in the trivial region, while the other wire lies in the nontrivial region. All three cases were studied before and some special properties were shown \cite{pablo, Zaz1, Jay2, Zaz2}, but a systematic comparison has not been given. Therefore, it is desirable to study the effect of the variation of the chemical potential in the wire. This would facilitate the experiment to tune the right region. Fig. 5(a) shows the contour plot for transient current as a function of the left wire's chemical potential $\mu_L$ and time period with a fixed step voltage $V=0.1\Delta$.  Here, we fixed $\mu_R=-2t_0$ in the nontrivial region. The period is $4\pi$ when both chemical potentials are in the nontrivial region, while it becomes $2\pi$ when the left wire lies in the trivial region.  Conversely, if the chemical potential of the right wire is set to $\mu_R=-2t_0+3\Delta$ within the trivial region, the period is always $2\pi$ as shown in Fig. 5(b).

The period is distinct between the pure non-trivial junction case and two other cases. It would be important to find out whether we can distinguish the other two cases through supercurrent.  Previous research revealed that the Josephson current is strongly suppressed for
an S wave-P wave Josephson junction (trivial-nontrivial junction) \cite{Zaz1}.  A suppression can indeed be observed in Fig. 5(a) and Fig. 5(c). However, this observation is not fully consistent with the previous literature.
This inconsistency arises because a single semiconductor superconducting wire preserves  both S and P wave parings, and the junction can not be seen as a pure S-P Josephson junction in the trivial-nontrivial case instead of a blend between an S-S and an S-P Josephson junction. As a result, the strong suppression is weakened.
Another feature is that the $2\pi$ periodicity current is stable when both wires are in the trivial region, while it is unstable when only one wire is in the nontrivial region. Such a feature can be seen more clearly in the frequency map. Fig. 5(b) and Fig. 5(d) show the Fourier transformation of Fig. 5(a) and Fig. 5(c), respectively. The frequencies are mainly distributed around $f_0=2eV/h$ when both wires lie in the trivial region.
However, an enhanced higher order frequency $2f_0 = 4eV/h$ emerges when one wire enters into the nontrivial region.
This enhanced higher order Josephson current is caused by the finite coupling of MQPs.  Two MQPs together
can favor the $2f_0$ Josephson current \cite{qi1}.  Recent experiment with superconductor-TI-superconductor system also found
an enhanced $2f_0$ in the nontrivial region \cite{Erw2}, while the signal of $2f_0$ is small when the system lies in the trivial region. Our numerical
results are consistent with the experiment. These properties can be used to distinguish which region the system lies in, and they indicate how the chemical potential can be tuned under different conditions.

\subsection{Detrimental disorder in a Josephson current}  ~\label{sec:3}
 \begin{figure}[ptb]
\includegraphics[width=1.0\columnwidth]{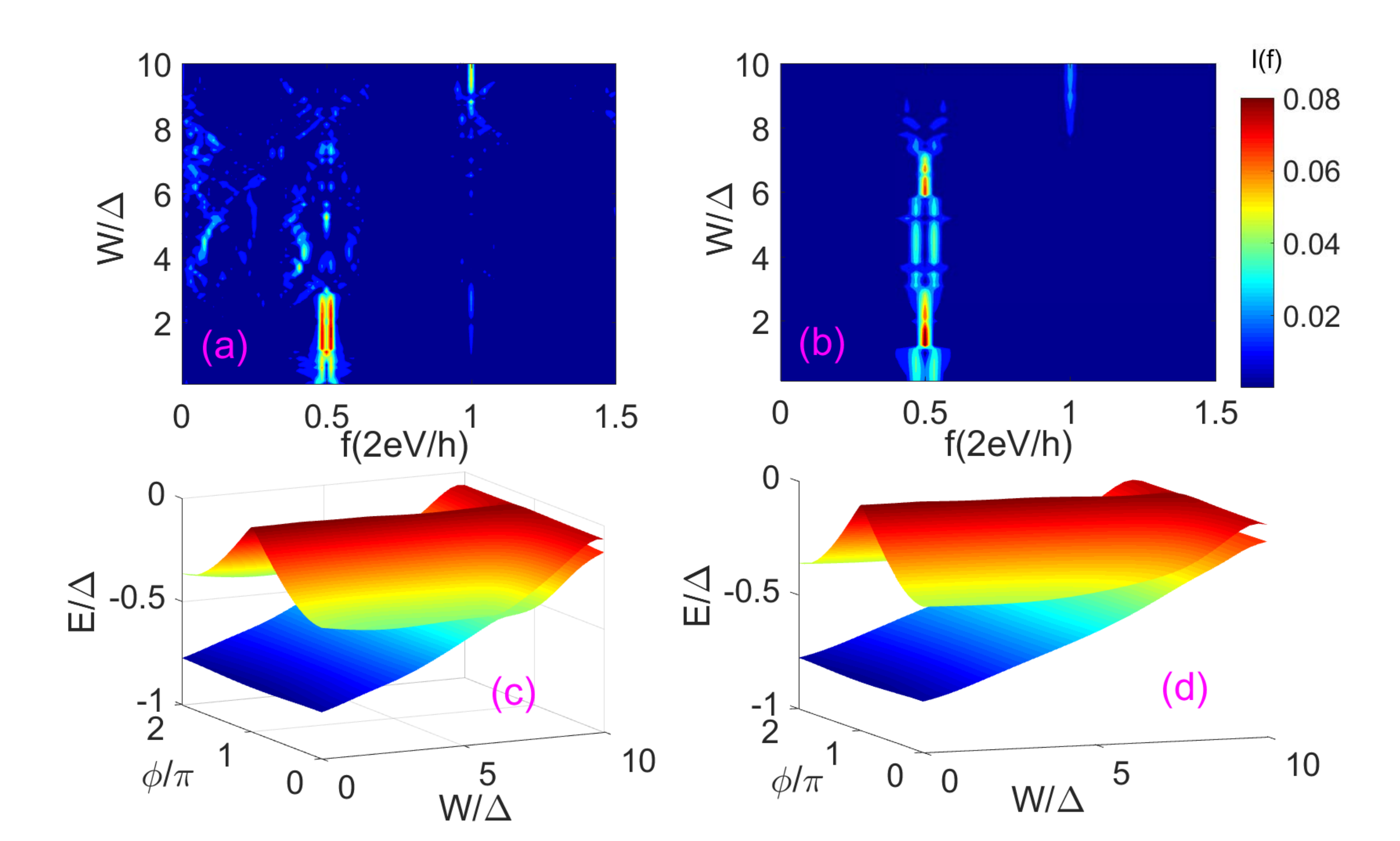}
\caption{(Color online) Adverse disorder in Josephson current. (a)Fourier transformation of transient current as a function of disorder strength and frequency.  Here disorder is randomly distributed at both wires. The
fractional Josephson effect is destroyed in weak disorder. (b)Fourier transformation of transient current as a function of disorder strength and frequency. Here disorder is zero around the junction and other positions are the same as (a). In this case, the fractional Josephson effect can sustain in strong disorder. (c)The spectra of Andreev bound states in (a). (d) The spectra of Andreev bound states  in (b), as a function of disorder and phase difference.   Other parameters are: $\mu_L=\mu_R = -2t_0, V=0.1\Delta$.}\label{Figure5}%
\end{figure}
One of the advantages of numerical simulation is that the effect of disorder can be investigated. In this section, we study the effect of disorder on the dynamic evolution of the Josephson current.  Fig. 6(a) shows the frequency map as a function of disorder strength and frequency. The disorder here is an on-site one, which is uniformly distributed over the range [-W/2,W/2].  The fractional Josephson current is modified with very small disorder. However, a sudden destruction occurs at about $W=3\Delta$.  In a single topological superconducting wire, the MQPs can survive upon the disorder strength on the order of $t_0$ \cite{jie}. Comparing to the robust MQPs, the fractional Josephson current is very fragile. The reason for the detrimental disorder effect in a Josephson current needs to be clarified. We found that it
is mainly due to the collapse of the bulk band. Fig. 6(c) shows the evolution of energy bands as a function of disorder and phase difference.
the Andreev bound states formed by two MQPs are robust against disorder, and
they are not destroyed until $W=8\Delta$. However, the
second band collapses soon with the effect of disorder. As we pointed out in previous sections, sustaining a $4\pi$ Josephson current requires $J_1+V<\Delta_{eff}$. Thus, the fractional Josephson current is sensitive to disorder.  Hence, we can conclude:
 Experimenters should be very careful  with respect to  the effect of disorder on bulk bands when combine more topological superconducting wires together. The dynamical non-Abelian braiding process should keep the MQPs far away from the bulk states. The disorder-induced collapse of bulk bands has unexpected implications for the topological quantum computation: the topologically protected long dephasing time in topological quantum computing would be largely suppressed by the collapse of bulk bands.
 Recently, the research groups led by Kouwenhoven and Marcus fabricated the semiconducting superconducting junctions, but failed to observe
 the $4\pi$ Josephson current \cite{kou1,nkou1,nkou2,nMar1}. The detrimental effect of disorder may help explain this. To address this problem, Fig. 6 (b) shows
 the frequency map as a function of disorder strength and frequency.  Here, all parameters are the same as in Fig. 6(a), except that disorder
 at the junction is eliminated. We set the disorder strength to zero for the 6 sites around the junction. Interestingly, the fractional Josephson effect
 can be functional up to $W=8\Delta$. Fig. 6(d) shows the energy bands as a function of disorder strength and phase difference. Although the bulk bands
 still collapse due to the effect of disorder, a safe gap always exist between the topological Andreev bound states and the bulk bands.
 Thus, reducing the interface disorder can reduce the detrimental effect of disorder.

\section{Conclusions}
~\label{sec:conclusions}
In summary, we studied numerically the transient transport properties of a topological Josephson junction, demonstrating that a $4\pi$ transient current can be observed  under
the suitable conditions. The properties of the transient current were then investigated under different conditions.
In the first case, both topological superconductor wires are in the topologically nontrivial region, which leads to the observation of a $4\pi$ periodic Josephson current. In the second case, when one wire is in the trivial region, while the other is in the nontrivial region, an enhanced $\pi$ periodic Josephson current is formed. In the third case, both topological superconductor wires are in the trivial region, which shows a stable $2\pi$ periodic Josephson current. Analyzing the current properties for different conditions can help develop experiments to extract an actual $4\pi$ periodic Josephson current by tuning proper parameters. Furthermore, we investigated the effect of disorder at the junction and showed that the $4\pi$ periodic Josephson current is sensitive to disorder. Experimenters should design the system carefully to eliminate the effect of disorder. These results are useful for the experimental detection of the $4\pi$ periodic Josephson current.

Finally,  we want to discuss the observability of a $4\pi$ Josephson current in an experiment. The experimental values of the induced superconducting gaps are usually of the order of several hundred $\mu eV$, which corresponds to the $GHz$ frequency range. Such a frequency can be observed using rf techniques. Two recent experiments have observed the signal of Josephson emission using such techniques \cite{Deacon3,JR}.
  We expect that the observation of the fractional Josephson current will come soon in a semiconductor superconducting  wire.

\section*{ACKNOWLEDGMENTS}
We gratefully acknowledge the
support from NSF-China under Grant Nos. 11574245,  the China Post-doctoral Science Foundation Under Grant No. 2015M580828.

\end{document}